\documentclass[journal,comsoc]{IEEEtran}

\usepackage[style=ieee, mincitenames=1, maxcitenames=1, backend=biber]{biblatex}
\usepackage[pdftex]{graphicx}
\graphicspath{{img/pdf/}{img/png/}}
\usepackage{amsmath}
\usepackage{dsfont}
\usepackage[cmintegrals]{newtxmath}
\usepackage{bm}
\usepackage{stfloats}
\usepackage[caption=false]{subfig}
\usepackage{algorithm}
\usepackage{algpseudocode}
\usepackage{setspace}
\usepackage{comment}
\makeatletter
\newcommand{\removelatexerror}{\let\@latex@error\@gobble}
\makeatother
\usepackage{fixltx2e}
\usepackage{hyperref}

\usepackage{acronym}
\usepackage{tabularx}
\usepackage{booktabs}
\usepackage{multirow}
\usepackage{tikz}
\usetikzlibrary{matrix}
\usetikzlibrary{arrows,chains,matrix,positioning,scopes}
\usetikzlibrary{positioning}
\usetikzlibrary{shapes}
\usetikzlibrary{backgrounds}
\usetikzlibrary{calc}
\usetikzlibrary{shapes}
\definecolor{cat1}{HTML}{66c2a5}
\definecolor{cat2}{HTML}{fc8d62}
\definecolor{cat3}{HTML}{8da0cb}
\definecolor{cat4}{HTML}{e78ac3}
\definecolor{cat5}{HTML}{a6d854}
\definecolor{cat6}{HTML}{ffd92f}
\definecolor{cat7}{HTML}{d9d9d9}
\definecolor{cat11}{HTML}{66c2a5}
\definecolor{cat12}{HTML}{fc8d62}
\definecolor{cat13}{HTML}{8da0cb}

\makeatletter
\tikzset{
    database/.style={
        path picture={
            \draw (0, 1.5*\database@segmentheight) circle [x radius=\database@radius,y radius=\database@aspectratio*\database@radius];
            \draw (-\database@radius,1.5*\database@segmentheight) -- ++(0,-3*\database@segmentheight) arc [start angle=180,end angle=360,x radius=\database@radius, y radius=\database@aspectratio*\database@radius] -- ++(0,3*\database@segmentheight);
        },
        minimum width=2*\database@radius + \pgflinewidth,
        minimum height=3*\database@segmentheight + 2*\database@aspectratio*\database@radius + \pgflinewidth,
    },
    database segment height/.store in=\database@segmentheight,
    database radius/.store in=\database@radius,
    database aspect ratio/.store in=\database@aspectratio,
    database segment height=0.1cm,
    database radius=0.25cm,
    database aspect ratio=0.35,
}
\makeatother

\definecolor{encoderc}{HTML}{c5bedf}
\definecolor{corec}{HTML}{fdf9c0}
\definecolor{decoderc}{HTML}{c5bedf}
\definecolor{lookupc}{HTML}{cce7cf}

\newcommand{\mmd}{\mathop{\mathrm{MMD}}}
\newcommand{\mean}{\mathop{E}}

\newcommand{\eg}{\textit{e.g.}}
\newcommand{\ie}{\textit{i.e.}}
\newcommand{\modelname}{Deep-Generative Graphs for the Internet}
\newcommand{\model}{DGGI}
\newcommand{\dsname}{Internet Graphs}
\newcommand{\cename}{Filtered Recurrent Multi-level}
\newcommand{\ce}{FRM}
\newcommand{\ds}{IGraphs}
\newcommand{\thetitle}{Data-driven Intra-Autonomous Systems Graph Generator}

\newcommand{\norm}[1]{\left\lVert#1\right\rVert}

\newcolumntype{b}{X}
\newcolumntype{s}{>{\hsize=.5\hsize}X}

\newcolumntype{x}[1]{>{\centering\let\newline\\\arraybackslash\hspace{0pt}}p{#1}}

\makeatletter
\if@twocolumn
    
\else
    
\fi
\makeatother

\algnewcommand\algorithmicforeach{\textbf{for each}}
\algdef{S}[FOR]{ForEach}[1]{\algorithmicforeach\ #1\ \algorithmicdo}

\begin{document}

\acrodef{DL}{deep learning}
\acrodef{ML}{machine learning}
\acrodef{\model}{\modelname}
\acrodef{BA}{Barábasi-Abert}
\acrodef{BB}{Bianco-Barábasi}
\acrodef{EBA}{Extended Barábasi-Abert}
\acrodef{DBA}{Dual Barábasi-Abert}
\acrodef{IAG}{Internet AS Graph}
\acrodef{AS}{Autonomous System}
\acrodef{BRITE}{Boston University Representative Internet Topology Generator}
\acrodef{ITDK}{Internet Topology Data Kit}
\acrodef{GNN}{Graph Neural Networks}
\acrodef{MMD}{Maximum Mean Discrepancy}
\acrodef{SICPS}{Simulates Internet graphs using the Core Periphery Structure}
\acrodef{SubNetG}{SubNetwork Generator}
\acrodef{S-BITE}{Structure-Based Internet Topology gEnerator}
\acrodef{RKHS}{reproducing kernel Hilbert space}
\acrodef{ADAM}{Adaptive Moment Estimation}
\acrodef{IRL}{Internet Research Lab}
\acrodef{CAIDA}{Center for Applied Internet Data Analysis}
\acrodef{CDF}{cumulative density function}
\acrodef{GRU}{Gate Recurrent Unit}
\acrodef{BFS}{Breadth First Search}
\acrodef{DGGM}{Deep Graph Generative Model}
\acrodef{\ds}{\dsname}
\acrodef{\ce}{\cename}
\acrodef{RBF}{Radial Basis Function}
\acrodef{MLE}{Maximum Likelihood Estimation}
\acrodef{DeepGMG}{Deep Generative Model of Graphs}

\title{\thetitle}

\author{Caio~V.~Dadauto,
    Nelson L. S. da Fonseca,
    and~Ricardo~da~S.~Torres \thanks{C. V. Dadauto 
        and N. L. S. da Fonseca are with the Institute of Computing, UNICAMP --
        University of Campinas, Brazil. e-mail:
        \{caio.dadauto,~nfonseca\}@ic.unicamp.br} \thanks{R.
        da S. Torres is with the Department of ICT and 
        Natural Sciences, NTNU -- Norwegian University of Science and Technologies,
        {\AA}lesund, Norway. e-mail: ricardo.torres@ntnu.no and Wageningen Data Competence Center, Wageningen, The Netherlands. email: ricardo.dasilvatorres@wur.nl}
}%

\markboth{Journal of \LaTeX\ Class Files,~Vol.~00, No.~0,
    June~2020} {Dadauto \MakeLowercase{\textit{et%
            al.}}:Network Routing based on Attention Deep Learning for Graphs}

\maketitle

\begin{abstract}
    Accurate modeling of realistic network topologies is essential for evaluating novel Internet solutions.
    Current topology generators, notably scale-free-based models, fail to capture multiple properties of intra-AS topologies. While scale-free networks encode node-degree distribution, they overlook crucial graph properties like betweenness, clustering, and assortativity. The limitations of existing generators pose challenges for training and evaluating deep learning models in communication networks, emphasizing the need for advanced topology generators encompassing diverse Internet topology characteristics.

    This paper introduces a novel deep-learning-based generator of synthetic graphs representing intra-autonomous in the Internet, named  \ac{\model}.
    It also presents  a novel massive dataset of real
    intra-AS graphs extracted from the project \ac{ITDK}, called \ac{\ds}\footnote{
        \ac{\ds} is public available through the following \hyperlink{https://drive.google.com/file/d/1ZwMlMz4lYZIp4BYxhccdyI3kYIQEDjSn/view?usp=sharing}{link}.
    }.
    It is shown that  \ac{\model}  creates synthetic graphs that accurately reproduce
    the properties of centrality, clustering, assortativity, and node degree. The \ac{\model} generator overperforms existing  Internet topology generators. On average, \ac{\model}   improves the \acf{MMD} metric $84.4\%$, $95.1\%$, $97.9\%$, and $94.7\%$ for assortativity, betweenness, clustering, and node degree, respectively.
\end{abstract}

\begin{IEEEkeywords}
    Machine learning, Graphs and networks, Internet Topology, Topology Generator
\end{IEEEkeywords}

\IEEEpeerreviewmaketitle%

\section{Introduction}\label{sec:introduction}
\IEEEPARstart{G}{enerating} graphs that represent realistic network topologies 
is crucial for modeling, as well as for assessing novel protocols and traffic control mechanisms for the Internet~\cite{labovitz2001impact,gunes2007importance}.
The growing employment of deep learning models, particularly \ac{GNN},
in various communications and network mechanisms and protocols 
necessitates extensive and 
diverse datasets~\cite{suarez2022graph,jiang2021graph,tang2021survey}.
The availability of such datasets  achieved by  using  \ac{GNN} 
can benefit investigations of novel traffic control mechanisms and protocol proposals 
~\cite{suarez2022graph,jiang2021graph,tang2021survey}.
However, topology generators that encode multiple properties of the Internet topology are still
unavailable~\cite{vazquez2002large,mahadevan2007orbis,giovanni2015s-bite,jiao2020structural,bakhshaliyev2020generation}.

Node connections in the Internet topology graphs are largely modeled by heavily tailed distribution~\cite{yook2002modeling}.
The \ac{BA}~\cite{barabasi1999emergence} algorithm  is the most popular one for generating scale-free networks
(\ie, networks for which  power-law distribution can model the node relations), and it has served as the basis for several topology generators~\cite{medina2001brite, moshiri2018dual,bianconi2001competition,albert2000topology}. 

Most Internet topology graph generators are structure-based and focus on inter-AS topologies~\cite{jiao2020structural,bakhshaliyev2020generation,giovanni2015s-bite,lappas2010simple,elmokashfi2010scalability,mahadevan2007orbis}.
Moreover, empirical observations are typically used to model the structure of the Internet as a hierarchical composition of graphs. This hierarchy is
based on various assumptions, such as 
scale-free, uniform growth, and function fitting for node degree distribution. However, these assumptions only partially capture the characteristics of subgraphs in the Internet topology since they 
rely on a network growth pattern ultimately based on a power-law distribution.
Structure-based generators have been designed to synthesize graphs of hundreds of thousands of nodes 
but  do not accurately reproduce the intra-AS graph properties,
such as betweenness, node degrees, clustering, and assortativity (Section~\ref{sec:evaluation}).

Classical generators based on scale-free models exhibit commendable characteristics,
such as elegant mathematical formulations and a high degree of overall flexibility.
Their proven utility and efficiency across various studies underscore the strength of
their well-established theoretical foundations~\cite{hood2015predicting,barford2009network,swenson2013simulating}.

Although scale-free networks can accurately model the  node-degree distribution of graphs,
generators based on power-law assumptions 
do not capture other relevant
graph properties~\cite{canbaz2018router}, 
such as betweenness, clustering, and assortativity. The replication of graph metrics is pivotal
in addressing issues within communication networks. Several solutions for real-world
network issues rely on graph metrics within their formulation. Notably, centrality metrics, such as
betweenness, have found application in optimizing computer networks, transportation networks, and
recommendation systems, among others~\cite{mendoncca2020approximating,feng2022intrinsic}.
Additionally, clustering metrics have been investigated
to understand network mechanisms, including routing protocols~\cite{latapy2008complex}. The
examination of network robustness has involved the utilization of clustering coefficients, such as in
assessing network attack tolerance~\cite{albert2000error}. Assortativity, another pivotal graph
metric, plays a crucial role in quantifying network growth, making its accurate reproduction a
fundamental characteristic for any generator. Consequently, assortativity is commonly employed in 
evaluating Internet-like graph
generators~\cite{mahadevan2007orbis,jiao2020structural,bakhshaliyev2020generation}.

On the other hand, data-driven solutions for various studies on Internet protocols and traffic control mechanisms have relied on trivial topologies. They are typically based on only a 
few samples of real networks or synthetic ones generated by \ac{BA}-based 
models~\cite{suarez2022graph,tang2021survey}.
However, the use of unrealistic topologies may produce misleading assessments of the effectiveness of the performance of new solutions~\cite{gunes2007measurement,hidaka2008modeling,albert2000error}.

This paper proposes an 
intra-AS graph generator based on
deep learning, named \acf{\model}. It accurately reproduces the centrality, clustering, assortativity, and node degree metrics of Internet graphs.
\ac{\model} allows the customization of synthetic graphs 
and can generate an arbitrary number of synthetic parameterized graphs.
To our knowledge, this is the first paper to propose an intra-AS
graph generator based on deep learning.

This paper also introduces a novel dataset of  $90,326$
intra-AS subgraphs extracted from the sets of
large intra-AS graphs (millions of nodes), named \ac{\ds}~\cite{igraphs}. It was collected from the project \ac{ITDK} conducted by the \ac{CAIDA}~\cite{caida2020itdk}.
Such extraction employs the \acf{\ce} algorithm, which was designed to capture the node agglutination patterns found in the Internet and ensure that 
the sizes of the subgraphs will be in a predefined range.
\ac{\ds} is especially useful for training \ac{\model}.  
Furthermore,  incorporating \ac{\ds} allows more extensive investigations on network protocols and mechanisms based on simulation and emulation.
The inclusion of $90,326$ provided graphs facilitates comprehensive network scenarios.   

The main contributions of this paper are:
\begin{itemize}
    \item the introduction of a novel generator (\ac{\model}) based on deep learning for the generation of intra-AS graphs that
    encodes not only the node degree distribution of training data but also their centrality, clustering, and assortativity;
    \item the presentation of a new dataset composed of real intra-AS graphs (\ac{\ds}) extracted from the project \ac{ITDK}.
\end{itemize}

Compared with existing generators for the Internet, \ac{\model}  improves the \acf{MMD} similarity index~\cite{gretton2006kernel,you2018graphrnn}, on average,  $84.4\%$, $95.1\%$, $97.9\%$, and $94.7\%$ for assortativity, betweenness, clustering, and node degree, respectively. 
In the worst case, \ac{\model} improves by $13.1\%$ for assortativity, and in the best case, $99.8\%$ for clustering.

This paper is organized as follows: Section~\ref{sec:related-works} presents related work focused on generators for Internet topology graphs; Section~\ref{sec:metrics} shows the graph metrics used for the assessment of the proposed solution; Section~\ref{sec:model-desc} introduces the \ac{\model}; Section~\ref{sec:caida-dataset} introduces the \ac{\ds}  dataset; Section~\ref{sec:evaluation} describes the evaluation procedures adopted to validate the \ac{\model} generator and discusses the obtained results. Section~\ref{sec:conclusion} points out some conclusions and directions for future work.

\section{Related Work}
\label{sec:related-works}

This section describes existing work related to graph generation to represent  Internet-like topologies.
Table~\ref{tab:related-works} summarizes the
characteristics of the referenced work and shows how the \ac{\model} differs from other
generators. The models are classified based on the applicability of their generated graphs for deep learning 
training. Generators classified as ``Suitable for DL'' (vide Table~\ref{tab:related-works})
can synthesize realistic graphs for an arbitrary number of nodes, i.e., they are not restricted to generating large graphs with hundreds of thousands of nodes.

Most  generators employ algorithms based on  scale-free networks and 
power-law node degree   distribution~\cite{albert2000topology,elmokashfi2010scalability,mahadevan2007orbis,lappas2010simple, medina2001brite,bianconi2001competition,moshiri2018dual,jiao2020structural,giovanni2015s-bite,bakhshaliyev2020generation}.
The models in~\cite{jiao2020structural,giovanni2015s-bite,bakhshaliyev2020generation,lappas2010simple}
aim at generating  graphs as a composition of subgraphs (structures), \eg,  AS nodes,  core, and  periphery.

\begin{table*}[!t]
  \centering
  \caption{Overview of related work.}\label{tab:related-works}
  \begin{tabularx}{\textwidth}{x{.14\textwidth}x{.15\textwidth}x{.2\textwidth}x{.08\textwidth}x{.08\textwidth}x{.2\textwidth}}
    \toprule[1.2pt]
    Reference & Technique & Requirements & View & Suitable for DL & Validation \\
    \toprule[1.2pt]
    SICPS \cite{jiao2020structural} & Multi-Structure Decomposition & Number of Nodes and Structure Statistical Properties & Inter-AS & No & First-Order Moments of Node Degree, Assortativity, and Clustering \\
    \midrule[.1pt]
    SubNetG \cite{bakhshaliyev2020generation} & Scale-free and Hierarchical Decomposition & Number of Nodes and Component Power Law Coefficients & Intra-AS & No & First-Order Moments of Node Degree and Clique Sizes \\ 
    \midrule[.1pt]
    S-BITE \cite{giovanni2015s-bite} & Scale-Free and Core-Periphery Decomposition & Number of Nodes and Core-Periphery Statistical Properties & Inter-AS & No & First-Order Moments of Node Degree, Clustering, Betweenness, Closeness, and Clique Size\\ 
    \midrule[.1pt]
    Jellyfish \cite{lappas2010simple} & Scale-Free and Ring Decomposition & Number of Nodes and Ring Power Law Coefficients & Inter-AS & No & First-Order Moments of Node Degree\\
    \midrule[.1pt]
    IAG \cite{elmokashfi2010scalability} & Scale-free and Hierarchical Decomposition & Number of Nodes & Inter-AS & No & First-Order Moments of Node Degree\\
    \midrule[.1pt]
    Orbis \cite{mahadevan2007orbis} & $dK$-Series Preservation & Number of Nodes and $dK$-Series & Both & No & First-Order Moments of Node Degree and Betweenness\\
    \midrule[.1pt]
    BRITE \cite{medina2001brite} & Barábasi-Abert & Number of Nodes and Preferential Attachment Coefficients & Both & Yes & First-Order Moments of Node Degree\\
    \midrule[.1pt]
    DBA \cite{moshiri2018dual} & Dual Barábasi-Abert & Number of Nodes and Preferential Attachment Coefficients & $-$ & Yes & $-$\\
    \midrule[.1pt]
    BB \cite{bianconi2001competition} & Bianco-Barábasi & Number of Nodes and Preferential Attachment Coefficients & $-$ & Yes & First-Order Moments of Node Degree\\
    \midrule[.1pt]
    EBA \cite{albert2000topology} & Extended Barábasi-Abert & Number of Nodes and Preferential Attachment Coefficients & $-$ & Yes & First-Order Moments fof Node Degree\\
    \midrule[.1pt]
    \model~(our) & Deep Learning & Number of Nodes and DL Weights & $-$ & Yes & Multi-Order Moments of Node Degree, Clustering, Betweenness, and Assortativity\\
    \bottomrule
  \end{tabularx}
  \raggedright
  \begin{tabular}{r}$-$ : No applicable\end{tabular}
\end{table*}

The \ac{BA} algorithm~\cite{barabasi1999emergence} is based on the preferential
attachment property, \ie, nodes with the highest node degree values tend to have new links attached.
The \ac{BRITE}~\cite{medina2001brite} implements the \ac{BA} algorithm to generate Internet graphs
for inter-AS and intra-AS topologies. 

A set of generators based on the \ac{BA} algorithm has been proposed to  
introduce new strategies for the preferential attachment paradigm to enhance 
the ability to generate more realistic graphs. The \ac{EBA} algorithm randomly modifies
links beyond the preferential attachment~\cite{bianconi2001competition}.
The \ac{BB} algorithm introduces a set of parameters (fitness weights) to specialize the preferential
attachment mechanism~\cite{bianconi2001competition}. In contrast, the \ac{DBA} algorithm changes the
number of links to be attached to new nodes in a random way~\cite{moshiri2018dual}. 
The \ac{BB} algorithm can reproduce the empirical power-law decays for \ac{AS} graphs~\cite{vazquez2002large}, although
the generated graphs are general-purpose ones, \ie, they are not specific to 
Internet graphs.

Generators based on structure decomposition have been proposed to mitigate the limitations of \ac{BA}-based generators in reproducing Internet 
graph properties. 
The \ac{SICPS}~\cite{jiao2020structural} model partitions the Internet graphs into 16 structures
representing different statistical assumptions, including the power-law distribution.
The \ac{SubNetG}~\cite{bakhshaliyev2020generation} represents sub-networks and  routers
as a bipartite graph on the basis of their power-law distribution.
The \ac{S-BITE}~\cite{giovanni2015s-bite} and \ac{IAG}~\cite{elmokashfi2010scalability} generators  
use similar approaches to decompose the Internet into core and periphery, each with a different power-law distribution. 
The Jellyfish~\cite{lappas2010simple} generator captures the core (referred to as the ring) of the Internet topology, which relies on the assumption of a power-law distribution.

However, both structure-based generators and the \ac{BA}-based generators rely on the assumption
of power-law distribution, which restricts the generalization of node connectivity,
since the decay parameter of a power-law distribution is  insufficient to represent
the topology diversity in the Internet~\cite{yook2002modeling}. 

An alternative proposal consists of employing different structuring procedures. The Orbis generator~\cite{mahadevan2007orbis}, for example, creates Internet graphs by adopting 
the $dK$-distributions as a criterion to maintain the correlation node degree of subgraphs of size $d$.
Orbis uses $1K$ and $2K$ distributions, which refer to the node degree and the joint node degree 
distributions, respectively. Although the $dK$-distributions attempt to unify a wide
range of graph metrics~\cite{mahadevan2007orbis}, they focus only on the node degree, which can lead to a poor representation of other graph properties, such as clique formation and node centrality.

However, all of these aforementioned generators have been validated only by inspecting the 
first-order moments of the selected metrics (as indicated in Table~\ref{tab:related-works}).
Moreover, classical generators are often manually parameterized based on
specific methodologies involving the inspection of Internet topologies.
This process is time-consuming and demands a substantial manual overhaul of
the model parameters. Orbis is the only exception since it implements an automatic procedure to adapt to the considered network scenario.

In contrast, \ac{\model} generators do not
depend on power-law assumption, and the modeling is not focused only on the node degree.
Since \ac{\model} is a generator based on \ac{DL}. Thus, it can be trained for various network scenarios without substantial overhaul.
Furthermore, our evaluation is not restricted to
the first-order moments of graph properties; we explore higher-order moments using
the \ac{MMD} metric to quantify the similarity between distributions of graphs.

\section{Graph Metrics}
\label{sec:metrics}

Four graph metrics are utilized to analyze of the properties of the generated graphs: node degree, coefficients of clustering, betweenness,
and assortativity~\cite{costa2007characterization}.

Let $G$ be an undirected graph with $N$ nodes, and $A\in\{0,1\}^N$ be the adjacent matrix of the graph $G$.
The node degree is the number of connections of a node, \ie, the node degree of the $i$-th node is
\begin{align}\label{eq:degree}
    \kappa_i = \sum_{j}^{N}A[i,j]\in\mathbb{N},
\end{align}
in which $A[i, j]$ is the element on the $i$-th row and $j$-th column of the adjacent matrix.

The clustering metric indicates the tendency of a node to cluster with its neighbors, \ie,
the occurrence of density-connected regions in the graph. The
clustering coefficient for the $i$-th node is defined as
\begin{align}\label{eq:clustering}
    \mathcal{C}_i = 2\frac{l_i}{\kappa_i(\kappa_i - 1)}\in[0, 1]
\end{align}
in which $l_i$ is the number of edges between the neighbors of node $i$.
Moreover, the global clustering coefficient can also be defined as 
\begin{align}\label{eq:globalclustering}
    \mathcal{C} = \frac{N_{\vartriangle}}{N_3}\in[0, 1]
\end{align}
in which $N_{\vartriangle}$ is the number of triangles in the graph, and $N_3$ is the number of connected
triads of nodes, \ie, all sets of three nodes that are connected by either two or three undirected edges.

The betweenness metric indicates the importance of a node in relation to the number of shortest paths that pass
through it. Since hubs usually have a central role in graphs, nodes in a hub tend to present
a larger betweenness coefficient. Formally, the betweenness for the $i$-th node can be defined as
\begin{align}\label{eq:betweenness}
    \mathcal{B}_i = \sum_{\substack{p\neq i\neq q\in\{1,\cdots,N\}\\p\neq q}}\frac{\sigma_i(p,q)}{\sigma(p,q)}\in[0, 1]
\end{align}
in which $\sigma_i(p,q)$ is the number of shortest paths between the $p$-th and $q$-th nodes that pass through the  $i$-th
node, and $\sigma(p,q)$ is the total number of shortest paths between $p$ and $q$.

The assortativity metric indicates the preferential connectivity of nodes of different node degrees, \ie, whether or not the network growth follows the preferential attachment pattern.  Assortativity is a scalar
measure that is defined as
\begin{align}\label{eq:assortativity}
    r = \frac{\sum_ke_{kk}}{\sum_k\alpha_{k}\beta_{k}}\in[-1, 1],
\end{align}
where $k$ is the node degree of graph $G$, $\alpha_k=\sum_{k'}e_{kk'}$, $\beta_k=\sum_{k'}e{k'k}$, and
$e_{kk'}$ is the number of edges of nodes with degree $k$ and nodes with degree $k'$.

\section{\acf{\model} Model}
\label{sec:model-desc}

Figure~\ref{fig:dig} illustrates the three procedures used to instantiate the \ac{\model} generator:
the construction of the training set based on the application of the \ac{\ce} algorithm, the training of the deep-generative graph model, and the generation of synthetic graphs.

For the construction of the training set, the input is  
a set of samples from large intra-AS networks, with lower and upper bounds reflecting 
the number of nodes. The final training dataset contains only graphs with a given number of
nodes in the defined range. 

The \ac{DGGM} is then trained using the training graphs created during this first procedure. The final procedure
is then responsible for using the trained model to synthesize intra-AS graphs based on two parameters: an optional list that defines the number of nodes and the number of graph samples. 
When not specified, the range defined is that of the construction of the training set.

The implementation of each procedure is outlined next.

\begin{figure*}[!t]
    \centering
    \input{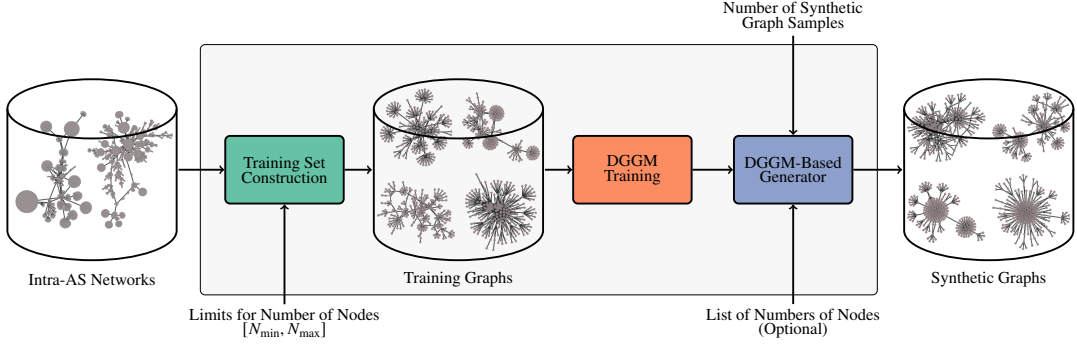}\label{fig:dig}
    \caption{
    The training pipeline for generative models tailored to the Deep Graph Generative Model (DGGM) is designed to synthesize Internet-like graph, which includes deriving a training dataset from limited samples of expansive networks and incorporating a layer for controlling both the quantity and number of nodes of the synthesized graphs. The instantiation of DGGI relies on three procedures: the training set construction, the DGGM training, and the synthetic graph generation based on DGGM.
    }~\label{fig:dig}
\end{figure*}

\subsection{Training Set Construction}
\label{sec:training-set-construction}

To extract subgraphs with a size bounded by a pre-defined limit,
we propose the \ac{\ce} algorithm, which employs the multi-level algorithm~\cite{blondel2008fast} recursively, followed by using  a filter based on betweenness centrality to avoid a single-star topology.

The multilevel algorithm aims to define the communities that maximize the
node's local contribution to the overall modularity score. This score measures
the ratio between the density of intra and  inter-community links. For a graph with $|E|$
edges, the modularity can be defined as 
\begin{align}
    Q = \sum\limits_{p_i\in\mathcal{P}}\left[\frac{S_{\mathrm{in}}^{p_i}}{2|E|}-\left(\frac{S_{\mathrm{tot}}^{p_i}}{2|E|}\right)^2\right]
\end{align}
where $\mathcal{P}$ is the set of graph communities, $p_i$ is the $i$-th community, $S_{\mathrm{in}}^{p_i}$ is the
number of edges into community $p_i$, and $S_{\mathrm{tot}}^{p_i}$ is the number of edges incident to the nodes in the community
$p_i$.

The multi-level algorithm operates in two steps. In Step 1, each node in a graph is assigned a distinct community.
For each node, it assesses the gain in modularity by relocating the node to the community of another node.
The node is placed in the community that yields the maximum positive modularity gain; if the gain is negative,
the node remains in its current community.

Step 2 involves constructing a new graph using the communities identified in Step 1 as nodes.
Nodes are formed by merging all nodes within a community into a single node. Links between nodes of the same community
result in self-loops for the community in the new graph.
Then, using this new graph, the algorithm iterates through Step 1 until no further improvement in modularity can be achieved.

The recursive application of the multi-level algorithm may lead to graphs with a single hub (graph with a star topology), due to the high disparity between intra and inter-hub links observed in subgraphs with a large number of hubs and low clustering coefficients.
In order to avoid the redundant occurrence of hubs in the training dataset, all subgraphs defined
by a single hub are discarded using the aforementioned filter.

\begin{figure}[!t]
    \vspace{-.5cm}
    \removelatexerror%
    \begin{algorithm}[H]
        \caption{\acf{\ce}}\label{alg:trainingset}
        \begin{algorithmic}[1]
            \Require~A graph $G$, the minimum ($N_{\min}$) and the maximum ($N_{\max}$) of the number of nodes.
            \State$\mathrm{to\_process} = [G]$
            \State$\mathrm{training\_set} = []$
            \While{$|\mathrm{to\_process}| > 0$}
                \State$G = \mathrm{to\_process}.pop()$
                \State$N = \mathrm{number\_of\_nodes}(G)$
                \If{$N > N_{\max}$}
                    \State{$\mathrm{clusters} = \mathrm{multilevel}(G)$}
                    \If{$|\mathrm{clusters}| > 1$}
                        \State{$\mathrm{to\_process} = \mathrm{concat}(\mathrm{to\_process}, \mathrm{clusters})$}
                    \EndIf
                \Else
                    \If{$N > N_{\min}$ and $|[\mathcal{B}_1,\cdots,\mathcal{B}_N] > 0| > 1$}
                        \State{$\mathrm{training\_set} = \mathrm{concat}(\mathrm{training\_set}, \mathrm{clusters})$}
                    \EndIf
                \EndIf
            \EndWhile%
            \State \Return~$\mathrm{training\_set}$
        \end{algorithmic}
    \end{algorithm}
    \vspace{-0.78cm}
\end{figure}

Algorithm~\ref{alg:trainingset} presents the \ac{\ce} algorithm used to construct the training set.
The lists defined in the first two lines control
the process of subgraph extraction. The list {\em \textrm{to\_process}} contains the graphs
that must be split into smaller subgraphs, while the list {\em \textrm{training\_set}} contains
the subgraphs that define the training set. 

 The first condition in Line 6 checks whether the graph in the loop has a node count
greater than the upper limit ($N_{max}$). If this condition holds, clusters are attempted to be
extracted from the graph under consideration using the multilevel algorithm, and the loop continues.
The condition in Line 8 ensures that the subgraphs are pushed to {\em \textrm{to\_process}} only
if the multilevel algorithm can split the graph. Otherwise, it is ignored.

Conversely, if the node count does not exceed $N_{max}$, the second condition in Line 12 comes
into play. This condition ensures that the graph under consideration meets the criteria of
having a node count greater than the lower limit ($N_{min}$) and is not structured as a
single star topology, \ie, a topology that has more than one node with a betweenness coefficient greater than zero.

The \ac{\ce} algorithm is a recursive algorithm designed to stop when all graph components fall
within the range of $[N_{\min}, N_{\max}]$. The best-case scenario occurs when a single call of the
multi-level algorithm successfully places all components within this range.
In this optimal case, the algorithm's complexity is  $\mathcal{O}(N)$ since the multi-level algorithm has a linear complexity
as a function of the number of nodes~\cite{blondel2008fast}.

Conversely, the worst-case scenario arises when each call of the multi-level algorithm
divides the original graph into two components, each having half the number of nodes of the original graph.
The termination condition for the \ac{\ce} algorithm is met when all graph components satisfy the node number constraints, \ie, be in the range of $[N_{\min}, N_{\max}]$.
In this worst-case scenario, the algorithm's complexity is determined by the structure of a binary tree,
where each leaf represents a graph component with nodes in the range $[N_{\min}, N_{\max}]$.
The complexity in the worst case is $\mathcal{O}\left(2(1-2^h)N\right)$, since the number of the  nodes in a binary
tree of height $h$ is the sum of a finite geometric series whose elements are the sum of nodes in each level of tree.
The tree height can be estimated as
\begin{align}
    &N_{\min} \le \frac{N}{2^h} \ge N_{\max} \\
    &\frac{N}{N_{\min}} \ge 2^{h} \le \frac{N}{N_{\max}} \\
    &\log_2\left(\frac{N}{N_{\min}}\right) \ge h \le \log_2\left(\frac{N}{N_{\max}}\right),
\end{align}
since the total number of leaves in a binary tree is $2^h$.

If the multi-level algorithm fails to generate clusters, \ac{\ce}
will stop the loop in Line 3 and return the training set, regardless of its content, due to the
condition in Line 8. Conversely, the convergence is guaranteed by the depletion of the stack
of graphs awaiting processing, a state achieved when either the multi-level algorithm
ceases to identify new clusters or all identified clusters contain
fewer nodes than $N_{\max}$.

\subsection{\acl{DGGM} Training}
\label{sec:dggmtraining}

\ac{\model} uses the recently proposed GraphRNN~\cite{you2018graphrnn} as the deep-generative graph model.
This model is a general-purpose data-driven generator of synthetic graphs based on deep learning.

GraphRNN is acknowledged as the current state-of-the-art method for generating graphs,
demonstrating the generation of synthetic graphs that exhibit greater realism compared to other
\acp{DGGM}, such as GraphVAE~\cite{simonovsky2018graphvae} and
\ac{DeepGMG}~\cite{li2018learning}.

The architecture of the GraphRNN comprises two different hierarchical Gate Recurrent Units (GRUs)~\cite{cho2014properties},
one to embed the graph representation
and the other to predict the new connections for each node.
\ac{GRU} is a variation of recurrent neural networks specialized in embedding relations established by long-ordered sequences of tensors into a state, a vector with  a pre-defined dimension (so-called latent dimension).
GraphRNN maps the graph generation into a sequential procedure based on edges added to each
node. In order to make this feasible, GraphRNN establishes a node order for processing all graphs
during the training. 

There is no unique node order to represent a graph using a sequence of nodes. The number of possible representations of such a graph based on sequences is $n!$ with $n$ being the number of nodes~\cite{you2018graphrnn}.
The pre-defined node order is used by GraphRNN to reduce the number of possible node permutations, thus improving the 
learning efficiency~\cite{you2018graphrnn}. GraphRNN uses the node order established by the \ac{BFS} algorithm
since different node permutations can be mapped onto a unique node order~\cite{you2018graphrnn}.


\begin{figure}[!t]
    \vspace{-.5cm}
    \removelatexerror%
    \begin{algorithm}[H]
        \caption{GraphRNN}\label{alg:graphrnn}
        \begin{algorithmic}[1]
            \Require~The maximum number of nodes $N_{\max}$, the latent dimension $L$, the transient dimension $M$, two GRU layers, $g: (\mathbb{R}^M, \mathbb{R}^L) \mapsto\mathbb{R}^L$ and $f: (\mathbb{R}, \mathbb{R}^L) \mapsto\mathbb{R}$, the initial  state $h_g$, and the start and end tokens, $SOS\in\mathbb{R}^M$ and $EOS\in\mathbb{R}^M$.
            \State$s_1=SOS$ and $i=1$
            \ForEach{$i\in\{1,\cdots,N_{\max}$}
                \State$h_g = g(s_i, h_g)$
                \State$s_{i+1}=s_i$
                \ForEach{$j \in \left\{1,\cdots,\min{(i-1, M)}\right\}$}
                    \State$h_f = f(s_{i + 1}[j], h_g)$
                    \State$s_{i + 1}[j] = h_f$
                \EndFor%
            \EndFor%
            \State \Return~$\{s_1, \cdots, s_{N_{\max}}\}$
        \end{algorithmic}
    \end{algorithm}
    \vspace{-.5cm}
\end{figure}

Formally, let $G$ be a graph with $N$ nodes defined by a given order.
A sequence to represent the graph $G$ is described as $s = \left(s_1, \cdots, s_N\right)$, with
$s_i\in\{0,1\}^{i - 1}$ being the vector representing all connection
between the node $i$ and the remaining $i - 1$ nodes.
Assuming the \ac{BFS} order for the nodes, the dimension of vector $s_i\;\forall i$ can be
bounded by a fixed number $M$~\cite{you2018graphrnn}, the transient dimension. Therefore, a vector $s_i\;\forall i\in\{2,\cdots,N\}$
can be redefined as
\begin{align}\label{eq:graphrnnstate}
    s_i = \left[A[\max(i, i - M), i], \cdots, A[i - 1, i]\right],
\end{align}
where $A[i, j]$ is the element of the $i$-th row and $j$-th column of the adjacency matrix of the graph $G$.

Algorithm~\ref{alg:graphrnn} outlines how GraphRNN synthesizes a graph. The loop in
Line 2  determines all the connections for each new node.
These connections are defined for all $N_{\max}$ nodes (as defined in 
Section~\ref{sec:training-set-construction}). In Line 3, the
state $h_g$ is determined by the \ac{GRU} $g$ using the connections of $i$-th node and the previous 
state $h_g$.
The loop in Line 5 determines the $\min(i - 1, M)$ connections of the $(i+1)$-th node,
in which $i$ nodes have already been added to the graph. Finally, each $j$ connection is determined by the second
\ac{GRU} $f$ using the current  $h_g$ state and
the $j$-th connection of the $i$-th node. The complexity of Algorithm~\ref{alg:graphrnn}
is $\mathcal{O}\left(\min(i - 1, M)N_{\max}\mathcal{C}_g + N_{\max} \mathcal{C}_f\right)$, where $\mathcal{C}_g$ and $\mathcal{C}_f$ are the respective complexities for GRU layers
$f$ and $g$.

\subsection{Synthetic Graph Generation}

The final procedure is responsible for synthesizing intra-AS graphs. It utilizes the trained GraphRNN model.
Two parameters are required to generate graphs: the number of synthetic graphs and, optionally, a list of the number of nodes.

Given a trained GraphRNN,  synthetic graphs are created using the output of Algorithm~\ref{alg:graphrnn}.
In order to define a graph, the vectors $\{s_1, \cdots, s_{N_{\max}}\}$ provided by the trained
GraphRNN is transformed into an adjacency matrix $A$.
This transformation requires the mapping of the vector values to be 0 or 1 since  $s_i\in{[0,1]}^M\;\forall i$.
A threshold $\tau$ is used to implement the mapping. The algorithm assigns 1 if the value is larger than $\tau$ and 0 
otherwise.

The use of a fixed value of $\tau$ will produce the same set of graphs for all
runs of Algorithm~\ref{alg:graphrnn} as a consequence of the fixed weights of GraphRNN.
To avoid such reproduction, the threshold $\tau$ is sampled from a uniform distribution 
$\mathcal{U}(]0,1[)$ each time the $M$ edges of a node are defined, which results in 
$N$ random parameters for each synthetic graph, with $N$ representing the number of nodes.

Given the statistical nature of both deep learning models and the threshold $\tau$, the adjacency matrix
resulting from the former process does not necessarily represent a connected graph. 
The connected subgraphs provided by the adjacency matrix provide the synthetic graph.

\begin{figure}[!t]
    \vspace{-.5cm}
    \removelatexerror%
    \begin{algorithm}[H]
        \caption{Graph Generator}\label{alg:graphgenerator}
        \begin{algorithmic}[1]
            \Require~A trained GraphRNN $\mathcal{F}$, the number of synthetic graphs $T$, and, optionally, the list of numbers of nodes $L$, the minimum ($N_{\min}$) and the maximum ($N_{\max}$) of the number of nodes, and the transient dimension $M$.
            \State$\mathrm{graphs} = []$
            \If{$L=\emptyset$}
                \State$L=\{N_{\min},\cdots,N_{\max}\}$
            \EndIf
            \While{$|graphs| < T$}
                \State$\tau\sim\mathcal{U}(]0,1[)$
                \State$A_k = \{0\}^{N_{\max}\times N_{\max}}$
                \State$\{s_1, \cdots, s_{N_{\max}}\} = \mathcal{F}()$
                \ForEach{$s_j\in\{s_1, \cdots, s_{N_{\max}}\}$}
                    \ForEach{$i\in\{\max\{1, j - M\}, j - 1\}$}
                        \If{$s_j[i] \ge \tau$}
                            \State$A_k[i, j] = 1$
                        \Else
                            \State$A_k[i, j] = 0$
                        \EndIf
                    \EndFor%
                \EndFor%
                \ForEach{$G = \mathrm{connected\_subgraphs}(A_k)$}
                    \If{$\mathrm{number\_of\_nodes}(G)\in L$}
                        \State$\mathrm{graphs}.\mathrm{push}(G)$
                    \EndIf%
                \EndFor%
            \EndWhile%
            \State\Return$graphs$
        \end{algorithmic}
    \end{algorithm}
    \vspace{-.5cm}
\end{figure}

Algorithm~\ref{alg:graphgenerator} outlines how the generator is defined using a pre-trained GraphRNN.
In Lines 6 to 8, the threshold, an empty adjacent matrix, and the node states are defined, respectively. 
The loop in Line 10 maps each vector state $s_j$ to the proper column of the adjacency matrix using the
threshold $\tau$ and the transient dimension $M$ (defined in Section~\ref{sec:dggmtraining}).
The loop in Line 16 extracts all connected subgraphs from the adjacency matrix. Only the
subgraphs with a certain number of nodes included in the list $L$ are considered. The algorithm stops when
the number of synthetic graphs is greater than or equal to $T$. The generation of each graph
has complexity bounded by $\mathcal{O}\left(\mathcal{C}_{\mathcal{F}} + N_{\max}^2\right)$ and $\mathcal{O}\left(\mathcal{C}_{\mathcal{F}} + N_{\max}^2 + M^2 - N_{\max}M\right)$, where $\mathcal{C}_\mathcal{F}$ is the complexity of GraphRNN.

If the list of node numbers $L$ includes values beyond the predetermined range
$[N_{\min}, N_{\max}]$ established by the training dataset, the convergence of
Algorithm~\ref{alg:graphgenerator} cannot be assured. This limitation arises due to
the specialization of the GraphRNN model, which was trained specifically
to produce graphs featuring a node count confined
within the specified bounds of $[N_{\min}, N_{\max}]$.

\section{The \acf{\ds} Dataset}
\label{sec:caida-dataset}

This section describes the construction of the proposed intra-AS graph dataset, \ac{\ds}.
It also discusses the properties of the graphs produced.

\subsection{Graph Construction}

The construction of \ac{\ds} follows the procedures described in
Section~\ref{sec:training-set-construction}.  The \ac{ITDK} repository
is used to extract graphs to compose \ac{\ds}.
\ac{ITDK} is a \ac{CAIDA} project comprising the router connectivity of a
cross-section of the Internet. \ac{ITDK}  stores the historical router-level
topologies, providing the IPv4/v6 traces,
the router-to-AS assignments, the router geographic location, and the DNS for all observed IP addresses.

The present procedure uses the IPv4 traces, the geographic locations, and  AS assignments. All this
information is structured in three different text files. 
The Internet cross-section provided by \ac{ITDK},
with more than 90 million nodes, is divided into \ac{AS} subgraphs using the router-to-\ac{AS}
tables and the IPV4 links. The links are pre-processed to extract the edges
since more than two nodes can share the same link, \ie, one link can have multiple edges.
Then, all edges with the predecessor and the successor within the same \ac{AS} are labeled as intra-AS edges. This procedure relies on the information provided by the router-to-AS table.

\begin{figure}[!t]
    \centering
    \includegraphics[width=\columnwidth]{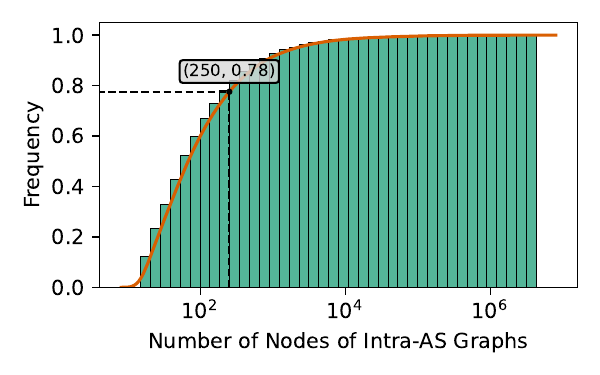}
    \caption{Cumulative distribution of the numbers of nodes of  intra-AS graphs.}\label{fig:asgraphdist}
\end{figure}

\begin{figure}[!t]
    \centering
    \includegraphics[width=\columnwidth]{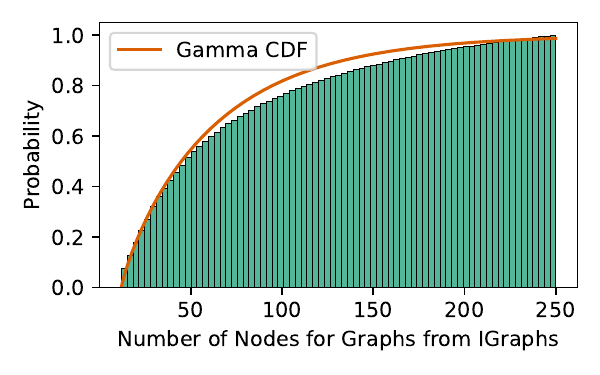}
    \caption{The cumulative distribution of the numbers of nodes of \ac{\ds}.
    The orange curve represents the \ac{CDF} of gamma with parameters  fitted to the presented
    cumulative distribution.}
    \label{fig:pdfcaidasize}
\end{figure}

The topologies are simplified to emphasize the router relations;
multi-edges (edges with the same pair of predecessor and successor) and
self-edges (edges with the predecessor equal to the successor) are discarded. Once
the intra-AS edges are defined, the graphs are created.

The number of graph nodes in \ac{\ds} ranges 
from $12$ to $250$. These limits have been defined based on the number of nodes of graphs often used to
evaluate recently proposed graph-based data-driven models in communication
networks~\cite{suarez2022graph,tang2021survey,bakhshaliyev2020generation}.
Nevertheless, the procedure described here can be used for
other ranges.

Figure~\ref{fig:asgraphdist} shows the distribution of the number of nodes of an \ac{AS} graph,
in which $22\%$ of the ASs have at least $250$  routers, a number that is out of the
range of interest. We also use the remaining set of graphs (those outside the specified range) can also be used, but after the application of the 
 procedures described in Section~\ref{sec:training-set-construction} to
extract the subgraphs that are indeed within the range of interest. The final dataset has $90,326$ graphs, each with 
$12$ to $250$ nodes. The distribution of the number of nodes is presented
in Figure~\ref{fig:pdfcaidasize}.

\begin{figure*}[!t]
    \centering
    \includegraphics[width=\textwidth]{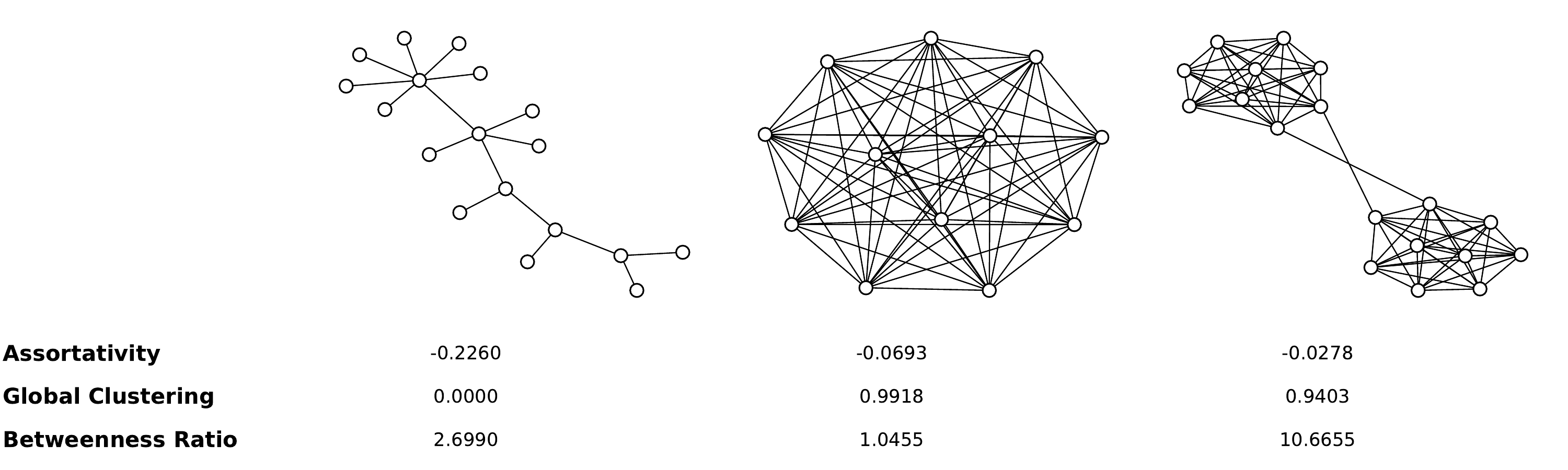}
    \caption{Examples of graph samples from \ac{\ds} presenting different graph metrics,
    evaluated in terms of their betweenness ratio, assortativity, and global clustering.
    The graph on the left does not contain triangles.
    This leads to a clustering coefficient equal to zero and a low assortativity score.
    The graph in the middle is a densely connected graph. In this case, the clustering coefficient is high, while the assortativity score is low. The graph on the right contains two hubs. In this case, the betweenness ratio score is high.}
    \label{fig:graphinstances}
\end{figure*}

The dataset also includes features associated with nodes and edges.
The geographic locations
of routers are used as node features, while the IP addresses of the links are used as the edge features.

\subsection{Analysis of Dataset Properties}

This section presents certain properties associated with the created graph dataset.



The analyses presented are based on three graph metrics: the coefficients for 
assortativity, clustering, and betweenness. For the latter, the ratio between the maximum and average betweenness coefficients. This ratio encodes the number of hop occurrences in each graph~\cite{knight2011internet}.
Figure~\ref{fig:graphinstances} illustrates the computed values associated with the three metrics
for three graph samples collected from \ac{\ds}.

\begin{figure*}[!t]
    \centering
    \subfloat[Assortativity]{\includegraphics[width=0.33\textwidth]{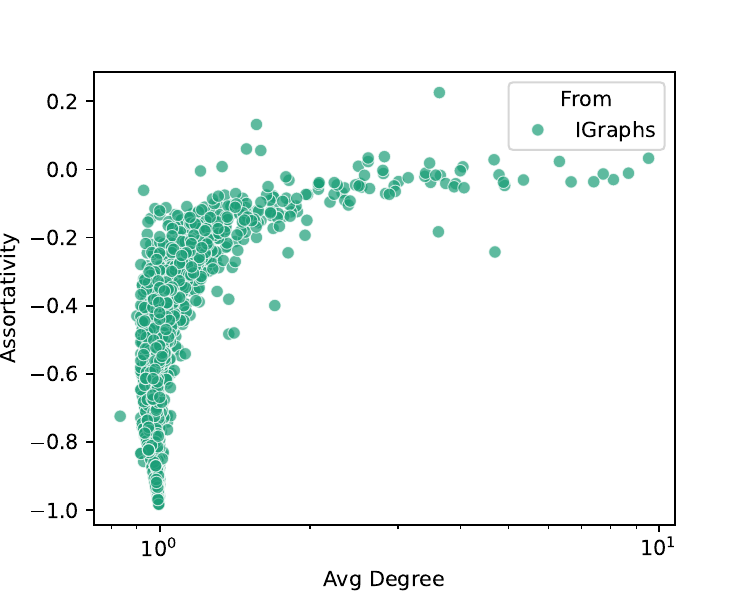}\label{fig:dataassortativity}}
    \hfil
    \subfloat[Global Clustering]{\includegraphics[width=0.33\textwidth]{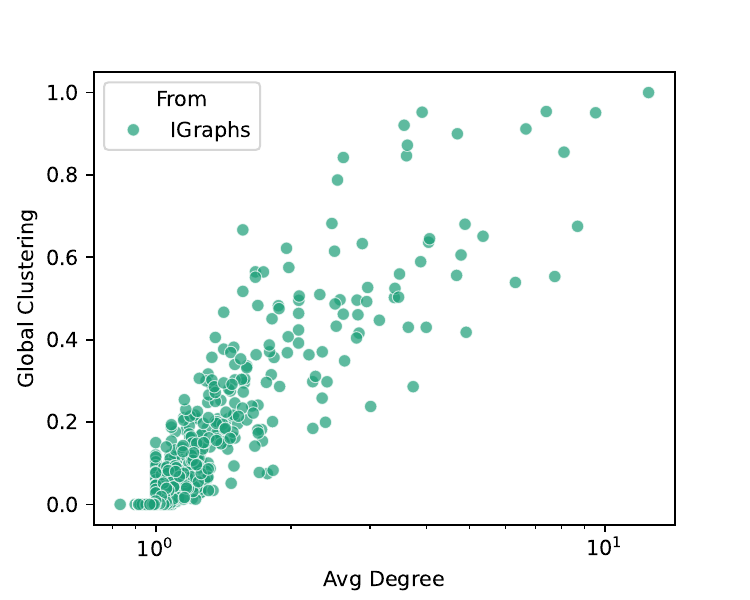}\label{fig:dataclustering}}
    \hfil
    \subfloat[Betweenness Ratio]{\includegraphics[width=0.33\textwidth]{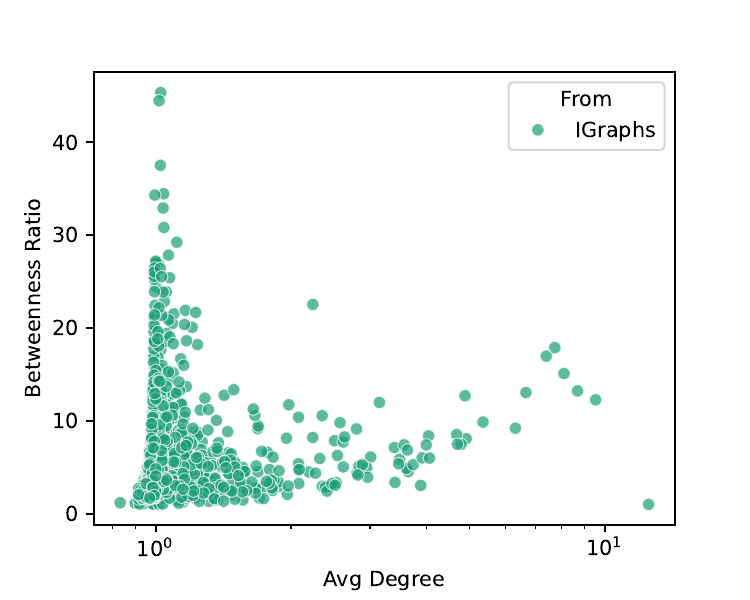}\label{fig:databetweenness}}
    \caption{The scattering of graph properties against the average node degree for \ac{\ds}  datasets. Figures (a) and (b) present the scattering of global assortativity and clustering, respectively; (c), shows  the scattering of the ratio between the maximum and the average of betweenness coefficients.}\label{fig:datascatter}
\end{figure*}


Figure~\ref{fig:datascatter} shows the scattering of the used graph metrics in relation to the average node degree
as computed by the approach adopted in~\cite{knight2011internet}.
The global clustering and the assortativity scores were computed as defined in
Equations~\eqref{eq:globalclustering} and~\eqref{eq:assortativity}.
The maximum and average values were computed according to Equation~\eqref{eq:betweenness} for the betweenness ratio.

\figurename~\ref{fig:dataassortativity} shows that the graphs from \ac{\ds}
with a larger average node degree tend to present null assortativity, and graphs with
nodes with a large node degree do not follow the precepts of preferential attachment, in contrast to graphs having 
small node degrees. In this case, the negative assortativity indicates an inverse preferential attachment
prone.

Figure~\ref{fig:dataclustering} shows the occurrences of density subgraph networks in \ac{\ds}. For graphs with a low average node degree, there is a trend to a monotonic
increase in global clustering, which is not 
verified for graphs with a high average node degree. These graphs present a greater dispersion, which
indicates that the nodes are not necessarily prone to triangle formation when the node degree increases.

\figurename~\ref{fig:databetweenness} suggests that a low
average node degree leads to a greater probability of hub occurrences. At the same time, the increase in
average node degree does not imply a large occurrence of hubs, \ie, the centrality
tends to be homogeneous among all the nodes.

\ac{\ds} comprises intra-AS graphs leveraging novel
possibilities for analyzing solutions in communication networks.
The $90,326$ graph samples from \ac{\ds} allow training
data-driven models based on real Internet topologies instead of augmentation procedures
over a few samples of real networks, \eg, from the Topology Zoo~\cite{knight2011internet} dataset.

Training using \ac{\ds} can also improve  the generalization capability, preventing over-fitting,
since \ac{\ds} presents large variability in topologies.

Typically,  the evaluations of network mechanisms are based either on  simulations or emulations running over simple topologies
(\eg, grids, stars) or on a few samples from real topologies (\eg, NSF, Geant2, and 
Germany50)~\cite{suarez2022graph,tang2021survey}. On the other hand, the introduction of \ac{\ds} opens opportunities for performing more robust
evaluations based on the diversity of real-world topologies.

\section{Evaluation of \acs{\model}}
\label{sec:evaluation}

This section describes the procedure followed to assess the effectiveness of the \ac{\model},
instantiated as described in Section~\ref{sec:model-desc}.
The assessment consists of quantifying to the extent generated synthetic topologies differ from real-world intra-AS graphs. 

Section~\ref{sec:mmd} introduces the  \acf{MMD} metric, which is used to estimate whether two samples
are modeled from the same distribution~\cite{gretton2006kernel}. Section~\ref{sec:training-procedure} describes the training procedures, while Section~\ref{sec:baselines} describes the baseline generators considered in our study. Finally, the results are presented and discussed in Section~\ref{sec:results-discussion}.

\subsection{\acf{MMD}}\label{sec:mmd}

Let $P$ and $Q$ be two distributions defined in a metric space
$X$, $\mathcal{H}$ be a \ac{RKHS} with a kernel $\kappa$, and
$\phi$ be a function that maps $X$ to $\mathcal{H}$. \ac{MMD} is defined as
\begin{align}
    &\mmd(P,Q) \coloneqq \norm{\mean_{x\sim P}\phi(x)-\mean_{y\sim Q}\phi(y)}_{\mathcal{H}}\nonumber\\
    &= \mathop{E}_{\substack{x\sim P \\ x'\sim P}}\kappa(x,x') - 2\mathop{E}_{\substack{x\sim P \\ y\sim Q}}\kappa(x,y) + \mathop{E}_{\substack{y\sim Q \\ y'\sim Q}}\kappa(y,y'),
\end{align}
which satisfies the metric properties, such as   $\mmd(P,Q) = 0$ iff $P = Q$~\cite{gretton2006kernel}.

The \ac{MMD} kernel used is defined as follows:
\begin{align}
    \kappa_{\mathcal{W}}(P, Q) =& \exp\left(\frac{\mathcal{W}(P, Q)}{\sigma^2}\right)
\end{align}
in which $\mathcal{W}$ is the Wasserstein distance and $\sigma$ is a free parameter
similar to a \ac{RBF} kernel. This function maintains
the \ac{MMD} assumptions since it induces a unique \ac{RKHS}~\cite{you2018graphrnn}, Proposition 2)
and all statistical moments, which can be verified by the Taylor expansion of
$\kappa_{\mathcal{W}}$~\cite{you2018graphrnn}.

The \ac{MMD} for each graph metric was estimated using bootstrap sampling given the computational cost of \ac{MMD} evaluation.
This procedure consists of sampling $500$ real graphs from the \ac{\ds} dataset with replacement and assessing the  \ac{MMD} using these real graphs with
other $500$ synthetic graphs created by either baselines or our \ac{\model} generator. This bootstrap evaluation was repeated $100$ times, allowing for the 
 establishment of a confidence interval associated with each \ac{MMD} value.

\subsection{Training the \model Generator}
\label{sec:training-procedure}

Our dataset, \ac{\ds}, was divided into three distinct parts: training, validation, and testing.
All sample graphs from \ac{\ds} were shuffled, and $70\%$  were reserved for the training set, while
the remaining graphs were divided equally for the validation and test sets.
This partition led to a total of $63.229$, $13.547$, and $13.547$ graphs for training,
validation, and testing, respectively.

The \ac{\model} generator learns the conditional distribution that models the link generation of the graphs in the training data, \ie,
given a previous state for graph representation, the generator predicts a new link. Binary cross entropy~\cite{topsoe2001bounds} is used as the loss function
since the prediction of links is mapped into a binary classification problem to determine if a link exists.

The machine used for training had an i7-9700 CPU  and a Quadro RTX 6000 GPU.
The training procedure consisted of $500$ epochs using the \ac{ADAM} optimizer~\cite{jlb2015adam} with an initial learning rate
of $0.003$, decaying by a factor of $0.3$ when the training reaches $300$ and $400$ epochs. Backpropagation 
for each mini-batch composed of $40$ graphs sampled from the training set was evaluated using uniform bootstrap sampling. The best
model was determined based on the best \ac{MMD} value obtained for the validation set, considering the node degree distribution.

\begin{figure*}[!t]
    \centering
    \subfloat[\ac{BA}-based baselines.]{\includegraphics[width=\columnwidth]{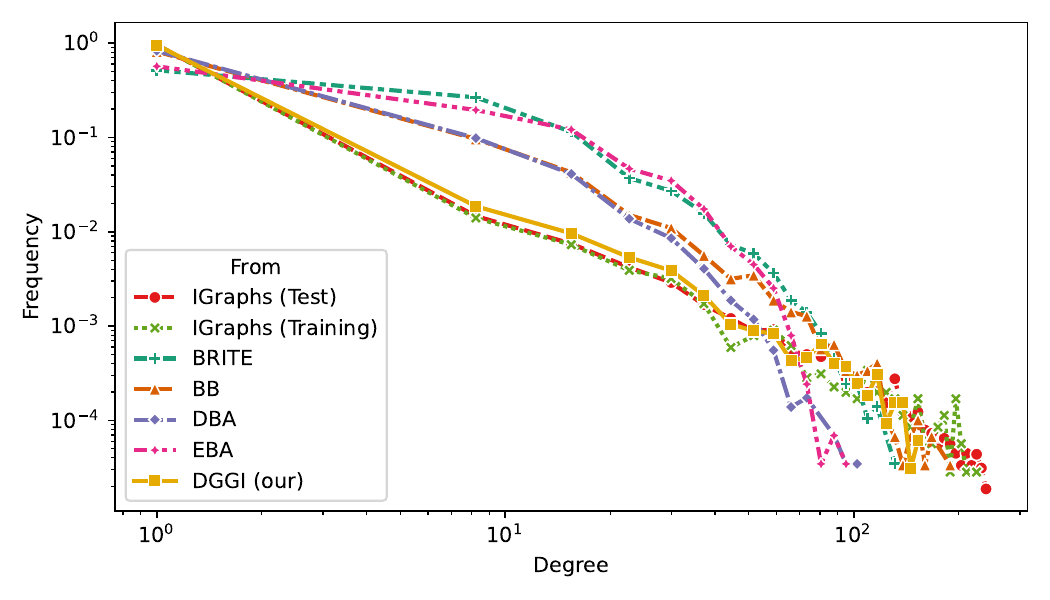}\label{fig:pdfbadegree}}
    \hfil
    \subfloat[Structure-based baselines.]{\includegraphics[width=\columnwidth]{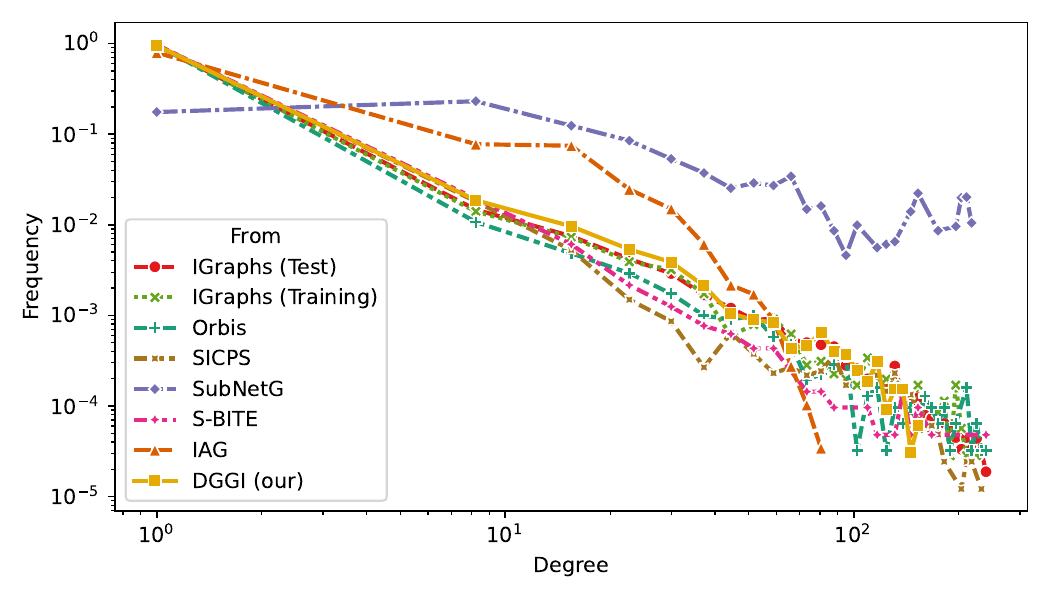}\label{fig:pdfnonbadegree}}\\
    \subfloat[\ac{BA}-based baselines.]{\includegraphics[width=\columnwidth]{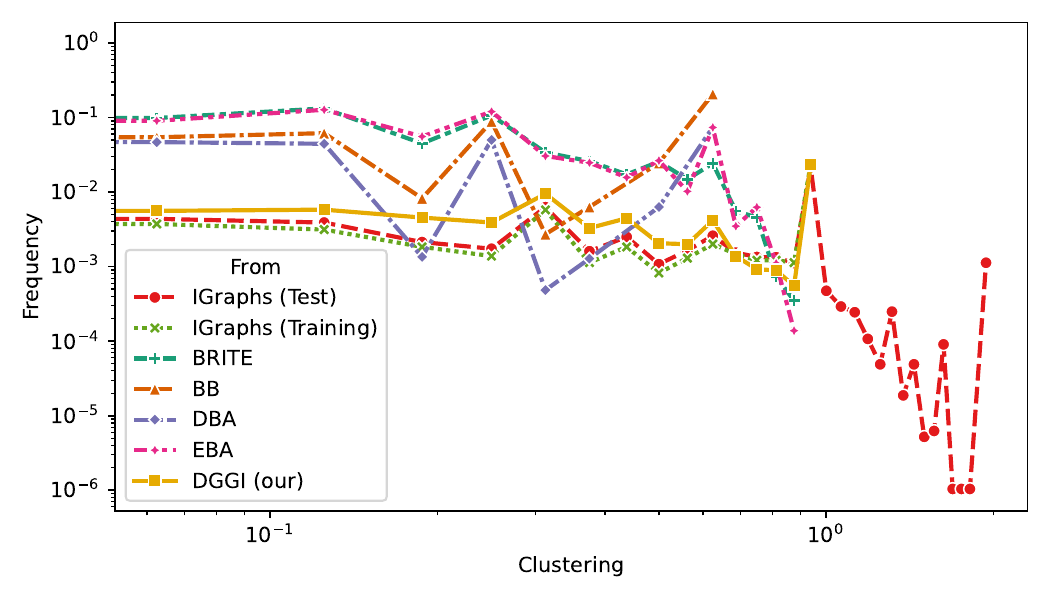}\label{fig:pdfbaclustering}}
    \hfil
    \subfloat[Structure-based baselines.]{\includegraphics[width=\columnwidth]{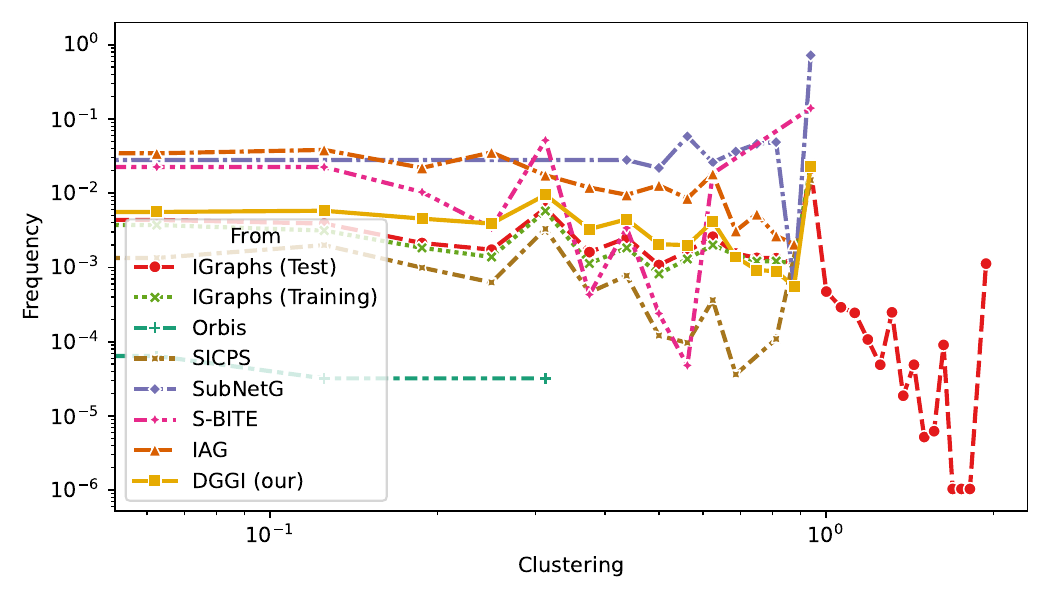}\label{fig:pdfnonbaclustering}}\\
    \subfloat[\ac{BA}-based baselines.]{\includegraphics[width=\columnwidth]{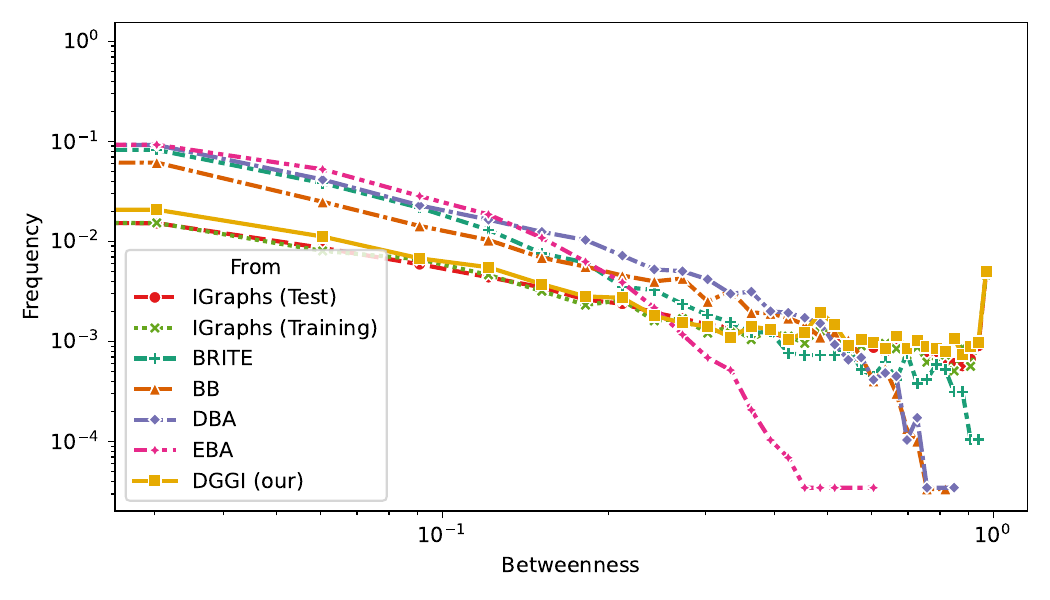}\label{fig:pdfbabetweeenness}}
    \hfil
    \subfloat[Structure-based baselines.]{\includegraphics[width=\columnwidth]{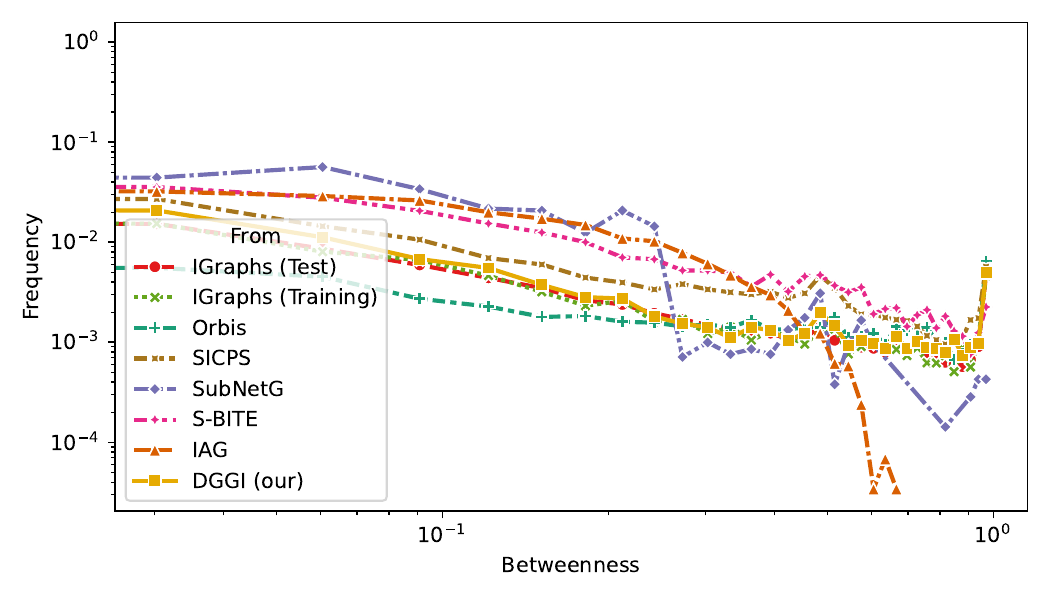}\label{fig:pdfnonbabetweenness}}\\
    \subfloat[\ac{BA}-based baselines.]{\includegraphics[width=\columnwidth]{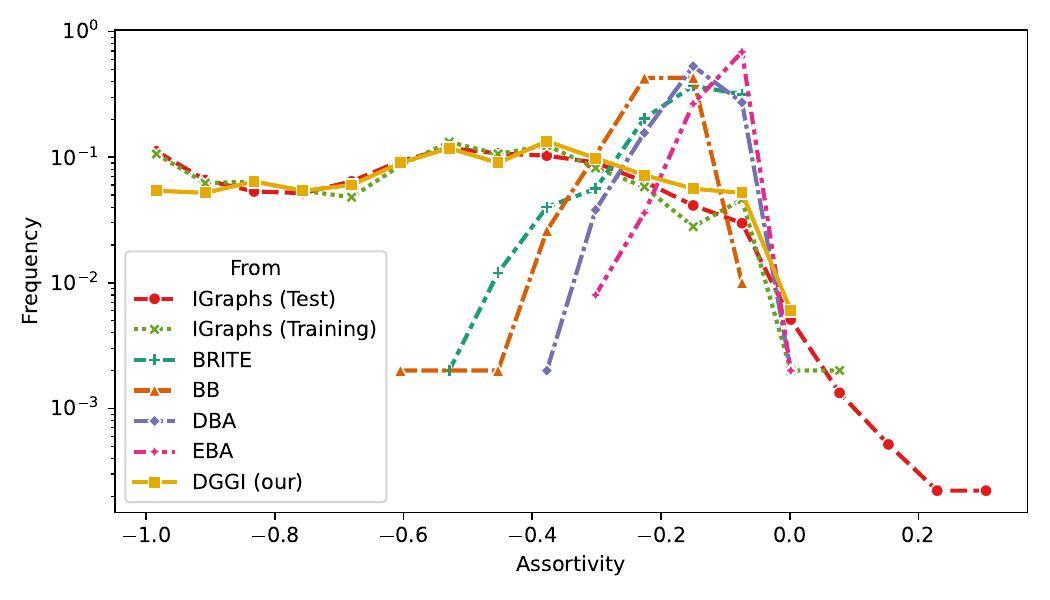}\label{fig:pdfbaassortativity}}
    \hfil
    \subfloat[Structure-based baselines.]{\includegraphics[width=\columnwidth]{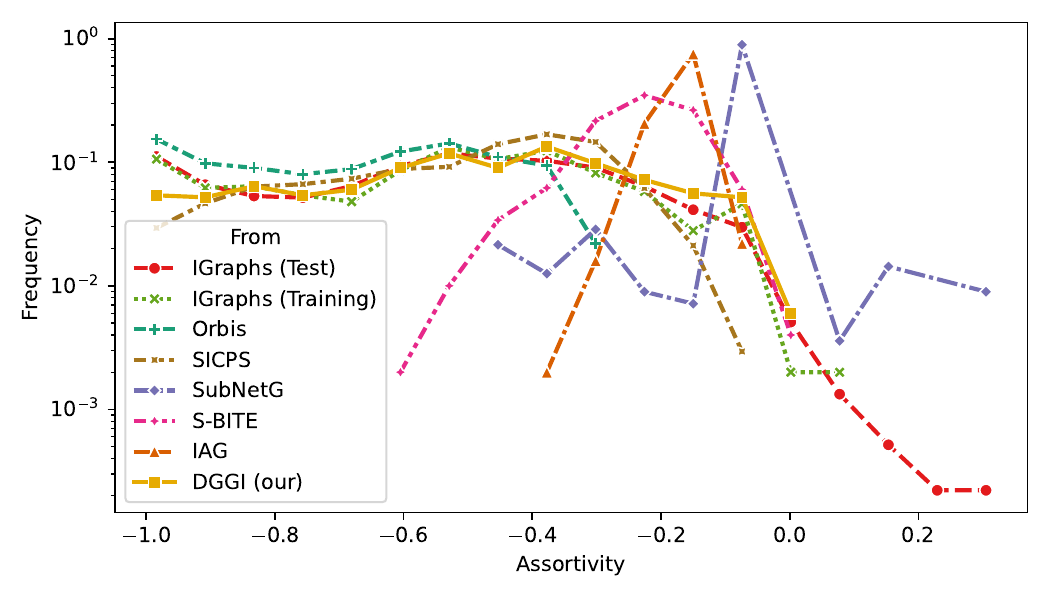}\label{fig:pdfnonbaassortativity}}
    \caption{The figure shows the frequency distribution for the occurrence of each assessed graph metric. For 
    comparison, all graphs present the distribution for CAIDA and for our model, \ac{\model}, and the baselines
    are organized in two groups, \ac{BA} and structure-based.
    }\label{fig:pdf}
\end{figure*}

\figurename~\ref{fig:pdf} shows the distributions of all mentioned graph metrics.
Based on these distributions, it can be inferred that the \ac{\model} generator reproduces the test set distribution accurately
for all assessed metrics.
Figures~\ref{fig:pdfbadegree},~\ref{fig:pdfbaclustering},~\ref{fig:pdfbabetweeenness}, and~\ref{fig:pdfbaassortativity} indicate that those  based on the \ac{BA} method (\ie, \ac{BRITE}, \ac{BB}, \ac{EBA}, \ac{DBA}) do not  reproduce the
 test set distribution for all metrics. 
 
\begin{table*}[!t]
  \centering
  \caption{Achieved average MMD for different graph metrics}\label{tab:mmd-values}
  \begin{tabular}{lcccc}
    \toprule[1.2pt]
    IGraphs&MMD Assortativity&MMD Betweenness&MMD Clustering &MMD Node Degree\\
    \toprule[1.2pt]
    Training Sampling
    &$7.09\mathrm{e}{-03}\pm 6.01\mathrm{e}{-03}$&$4.23\mathrm{e}{-04}\pm 3.96\mathrm{e}{-04}$&$7.15\mathrm{e}{-04}\pm 6.20\mathrm{e}{-04}$&$2.25\mathrm{e}{-03}\pm 1.59\mathrm{e}{-03}$\\
    \bottomrule
    \toprule[1.2pt]
    Generator&MMD Assortativity&MMD Betweenness&MMD Clustering &MMD Node Degree\\
    \toprule[1.2pt]
    BB&$1.58\mathrm{e}{+00}\pm 4.03\mathrm{e}{-02}$&$4.67\mathrm{e}{-02}\pm 2.25\mathrm{e}{-03}$&$9.62\mathrm{e}{-01}\pm 2.64\mathrm{e}{-02}$&$8.95\mathrm{e}{-01}\pm 1.26\mathrm{e}{-02}$\\
    \midrule[.1pt]
    BRITE&$1.70\mathrm{e}{+00}\pm 3.22\mathrm{e}{-02}$&$5.68\mathrm{e}{-02}\pm 2.57\mathrm{e}{-03}$&$4.54\mathrm{e}{-01}\pm 1.44\mathrm{e}{-02}$&$5.62\mathrm{e}{-01}\pm 1.26\mathrm{e}{-02}$\\
    \midrule[.1pt]
    DBA&$1.75\mathrm{e}{+00}\pm 3.28\mathrm{e}{-02}$&$7.99\mathrm{e}{-02}\pm 2.77\mathrm{e}{-03}$&$4.16\mathrm{e}{-01}\pm 2.32\mathrm{e}{-02}$&$1.06\mathrm{e}{+00}\pm 1.33\mathrm{e}{-02}$\\
    \midrule[.1pt]
    EBA&$1.85\mathrm{e}{+00}\pm 1.95\mathrm{e}{-02}$&$1.04\mathrm{e}{-01}\pm 4.16\mathrm{e}{-03}$&$8.54\mathrm{e}{-01}\pm 2.26\mathrm{e}{-02}$&$6.86\mathrm{e}{-01}\pm 1.26\mathrm{e}{-02}$\\
    \midrule[.1pt]
    IAG&$1.69\mathrm{e}{+00}\pm 3.36\mathrm{e}{-02}$&$7.87\mathrm{e}{-02}\pm 3.39\mathrm{e}{-03}$&$3.45\mathrm{e}{-01}\pm 1.75\mathrm{e}{-02}$&$8.51\mathrm{e}{-01}\pm 1.33\mathrm{e}{-02}$\\
    \midrule[.1pt]
    Orbis&$3.49\mathrm{e}{-01}\pm 5.63\mathrm{e}{-02}$&$4.59\mathrm{e}{-03}\pm 7.33\mathrm{e}{-04}$&$5.02\mathrm{e}{-02}\pm 7.29\mathrm{e}{-03}$&$2.35\mathrm{e}{-02}\pm 3.04\mathrm{e}{-03}$\\
    \midrule[.1pt]
    S-BITE&$1.45\mathrm{e}{+00}\pm 5.17\mathrm{e}{-02}$&$1.19\mathrm{e}{-01}\pm 8.64\mathrm{e}{-03}$&$6.24\mathrm{e}{-01}\pm 2.49\mathrm{e}{-02}$&$3.23\mathrm{e}{-01}\pm 1.69\mathrm{e}{-02}$\\
    \midrule[.1pt]
    SICPS&$8.70\mathrm{e}{-02}\pm 2.18\mathrm{e}{-02}$&$4.17\mathrm{e}{-02}\pm 5.50\mathrm{e}{-03}$&$2.76\mathrm{e}{-02}\pm 5.37\mathrm{e}{-03}$&$6.27\mathrm{e}{-02}\pm 9.18\mathrm{e}{-03}$\\
    \midrule[.1pt]
    SubNetG&$1.68\mathrm{e}{+00}\pm 3.66\mathrm{e}{-02}$&$1.03\mathrm{e}{-01}\pm 6.49\mathrm{e}{-03}$&$1.54\mathrm{e}{+00}\pm 1.76\mathrm{e}{-02}$&$5.95\mathrm{e}{-01}\pm 1.23\mathrm{e}{-02}$\\
    \midrule[.1pt]
    \ac{\model} (our)&$\mathbf{7.56\mathrm{e}{-02}\pm 3.26\mathrm{e}{-02}}$&$\mathbf{1.31\mathrm{e}{-03}\pm 8.77\mathrm{e}{-04}}$&$\mathbf{2.80\mathrm{e}{-03}\pm 1.54\mathrm{e}{-03}}$&$\mathbf{6.82\mathrm{e}{-03}\pm 2.17\mathrm{e}{-03}}$\\
    \bottomrule
  \end{tabular}
\end{table*}

\subsection{Baselines}
\label{sec:baselines}

The adopted baseline generators were divided into two groups: generators based on the \ac{BA} algorithm,
 \ac{BRITE}, \ac{BB}, \ac{EBA}, and
\ac{DBA}, and those based on the structure-based models of \ac{SubNetG},
\ac{S-BITE}, \ac{IAG}, and Orbis.

The \ac{BA}-based models were configured using commonly used parameter values~\cite{medina2001brite,yook2002modeling,vazquez2002large}.
Each new node establishes two new links following preferential attachment.
In \ac{BRITE}, these links connect the new nodes with existing ones. \ac{EBA}
was configured to behave as \ac{BRITE} $50\%$ of the time, while $25\%$ of the links were added to exiting nodes, and for the other $25\%$,
two known links were rewired. \ac{DBA} uses two \ac{BA} models simultaneously;  $35\%$ of the time, one link was added to
any new node, instead of two links.

The  number of nodes of the synthetic graphs generated by \ac{BA}-based models was
randomly determined. In order to improve the similarity between these graphs and \ac{\ds} graphs,
those generated by BA-based models
were adjusted to have a number of nodes drawn from a customized gamma distribution. This distribution was
tailored by fitting it to the distribution of the number of nodes of \ac{\ds} graphs, as illustrated in
\figurename~\ref{fig:pdfcaidasize}. The fitting process consisted in the employment of \ac{MLE},
resulting in a local optimal gamma distribution expressed with respect to $x$
as $(y^{a - 1}\exp{(-y)} / (s\Gamma(a))$, with $a = 0.852$, $s = 59.64$, $y = (x - m) / s$, and $m = 11.99$.

On the other hand, for structure-based models, the used configurations followed the parameters suggested in~\cite{jiao2020structural,giovanni2015s-bite,bakhshaliyev2020generation,mahadevan2007orbis}.
For \ac{S-BITE} and \ac{SubNetG}, the parameters were determined using topologies provided by
the \ac{IRL}-based dataset~\cite{giovanni2015s-bite,bakhshaliyev2020generation}. The joint distribution of node degrees required by Orbis was extracted
from the AS topology of CAIDA (Section~\ref{sec:caida-dataset}). Moreover, \ac{IAG} did not require any further configuration.

Unlike \ac{BA}-based models, \ac{S-BITE}, \ac{SubNetG}, and Orbis models were designed to generate graphs in the range of
hundreds of thousands of nodes. Since graphs with hundreds of nodes are expected, the \ac{\ce} algorithm
was used to extract small subgraphs from the large synthetic graphs provided by those baselines (Section~\ref{sec:caida-dataset}).

\subsection{Results and Discussion}
\label{sec:results-discussion}

We aim to assess the similarity of the graphs synthesized by the \ac{\model} generator 
considering the test set extracted from \ac{\ds} (Section~\ref{sec:training-procedure}). The same comparison was performed for the baseline generators. The similarity is quantified using the first and higher moments of the following four graph
metrics: node degree, clustering, betweenness, and assortativity.

Figure~\ref{fig:pdf} illustrates the distributions
for the considered graph metrics. It displays results for  two sets of baselines, \ac{BA}-based
and structure-based generators. To streamline the comparison, the distributions for the real graphs
(sampling from training and test datasets) and  the graph synthesized by \ac{\model} are also
provided.
However, the distributions in \figurename~\ref{fig:pdf}  allow only a visual comparison and give a
limited notion of mean, variance, and other higher
moments. Thus, the \ac{MMD} was used to represent the differences between the test
and training sets of these statistical moments in a concise form.

For structure-based baselines, Figures~\ref{fig:pdfnonbadegree},
\ref{fig:pdfnonbaclustering}, \ref{fig:pdfnonbabetweenness}, 
and \ref{fig:pdfnonbaassortativity} show that neither \ac{IAG} nor \ac{SubNetG} succeeded
in reproducing the  distributions of the test and training sets.
Orbis and \ac{SICPS} can visually decrease the overall distance concerning the 
distributions of the test and training sets for the node degree, betweenness, and assortativity.
However, Orbis does not reproduce the clustering distribution accurately.
accurately by Orbis. \ac{S-BITE} does not visually
reproduce the  right tail of 
distributions of the test and training sets for the clustering, betweenness, and assortativity metrics.

In contrast, Figures \ref{fig:pdfbadegree}, \ref{fig:pdfbaclustering}, \ref{fig:pdfbabetweeenness}, and
\ref{fig:pdfbaassortativity} depict the distributions for \ac{BA}-based baselines.
Regarding node degree, all \ac{BA}-based generators exhibit similar behavior, with baseline distributions
shifted relative to real ones (training and test sampling from IGraphs),
except for the distribution tails, which  We aim to assess the similarity of the graphs synthesized
are reasonably reproduced by
\ac{BA}-baselines. For betweenness, the behavior of the baselines mirrors that observed for node degree;
however, only \ac{BRITE} accurately reproduces the right tail of the distribution.
Lastly, \ac{BA} baselines prove inadequate in reproducing the
observed distributions, for clustering and assortativity metrics,.

Table~\ref{tab:mmd-values} shows the \ac{MMD} values assessed for the four graph metrics to quantify
the similarity between baseline distributions and the real distribution defined by the sampling from the test set.
Orbis and \ac{SICPS} outperform other baselines regarding \ac{MMD} values across all graph metrics,
suggesting their ability to replicate higher-order statistical properties of real intra-AS topologies.
Generally, baselines utilizing \ac{BA} model exhibit inferior performance in reproducing intra-AS topology,
as evidenced by their lower \ac{MMD} values compared to structure-based counterparts.
However, an exception is noted with \ac{BRITE},
despite its status as the earliest generator, as it surpasses all baselines except Orbis and \ac{SICPS}
in replicating the four metrics pertaining to high-order statistical properties.
Additionally, Table~\ref{tab:mmd-values} presents the \ac{MMD} values for training sampling,
which can serve as a reference due to the sampling of both training and test sets from the same population,
which is corroborated by the low levels of \ac{MMD} values observed in the training sampling.

\begin{table*}
    \centering
    \caption{Achieved average improvements using \ac{\model}}\label{tab:improvements}
    \begin{tabular}{lcccc}
    \toprule[1.2pt]
    IGraphs & Assortativity & Betweenness & Clustering &  Node Degree \\
    \toprule[1.2pt]
    Training Sampling & -967.05\% & -210.81\% & -291.14\% & -202.99\% \\
    \bottomrule
    \toprule[1.2pt]
    Baseline & Assortativity & Betweenness & Clustering &  Node Degree \\
    \toprule[1.2pt]
    BB       &        95.22\% &      97.19\% &     99.71\% &  99.24\% \\
    \midrule[.1pt]
    BRITE    &        95.56\% &      97.68\% &     99.38\% &  98.79\% \\
    \midrule[.1pt]
    DBA      &        95.68\% &      98.35\% &     99.33\% &  99.35\% \\
    \midrule[.1pt]
    EBA      &        95.91\% &      98.74\% &     99.67\% &  99.01\% \\
    \midrule[.1pt]
    IAG      &        95.54\% &      98.33\% &     99.19\% &   99.2\% \\
    \midrule[.1pt]
    Orbis    &        78.31\% &      71.35\% &     94.43\% &  70.93\% \\
    \midrule[.1pt]
    S-BITE   &        94.78\% &       98.9\% &     99.55\% &  97.89\% \\
    \midrule[.1pt]
    SICPS    &        13.08\% &      96.85\% &     89.86\% &  89.11\% \\
    \midrule[.1pt]
    SubNetG  &        95.49\% &      98.72\% &     99.82\% &  98.85\% \\
    \bottomrule
    \end{tabular}
\end{table*}

The node degree, indicative of the number of adjacent nodes, exhibits a notable correlation with betweenness centrality
owing to considering neighboring nodes in determining shortest paths. This correlation is depicted in
Figure~\ref{fig:pdf}, wherein the generators consistently manifest analogous errors in estimating node degree and
betweenness centrality. Notably, baseline methods, particularly Orbis, \ac{SICPS}, and \ac{BRITE}, tend to synthesize
network topologies exhibiting similar sets of shortest paths. The distribution of betweenness coefficients offers
insights into the prevalence of hub nodes within a topology. As illustrated in Figure~\ref{fig:pdfbabetweeenness},
\ac{BRITE}-synthesized graphs exhibit lower frequencies of large betweenness coefficients compared to real-world
counterparts from training and testing datasets, indicating a higher incidence of hubs in actual intra-Autonomous
System (AS) topologies. Conversely, Orbis and \ac{SICPS} demonstrate comparable hub frequencies to real-world
datasets, attributed to their ability to reproduce the right tail of the betweenness distribution.

Clustering, denoting the prevalence of triadic relationships among nodes, as outlined in Section~\ref{sec:metrics},
manifests in densely interconnected regions within a network. However, none of the baseline models successfully
reproduce the clustering coefficients observed in real intra-AS topologies, as evident from
Figures~\ref{fig:pdfbaclustering} and~\ref{fig:pdfnonbaclustering}. Furthermore, Table~\ref{tab:mmd-values} reports 
clustering \ac{MMD} values significantly deviating from zero for baseline methods, indicating a substantial
dissimilarity with real-world clustering patterns, up to 20 to 30 times worse concerning  clustering \ac{MMD}
values for training sampling. Thus, networks generated by baseline methods tend to exhibit dense node regions inconsistent with real intra-AS topologies. 

Assortativity, indicating the correlation between connections of nodes sharing similar neighborhood sizes,
remains unattainable for most baseline models, with Orbis and \ac{SICPS} being exceptions.
As illustrated in Figure~\ref{fig:pdf} and Table~\ref{tab:mmd-values}, baseline models fail to replicate the
assortativity coefficients observed in real intra-AS graphs, implying a divergence from the attachment tendencies
characteristic of actual network formations. Conversely, Orbis and \ac{SICPS} models demonstrate a reasonable approximation of this attachment tendency, aligning more closely with observed network assortativity patterns.

\ac{\model} outperforms all baselines regarding the realism of its synthetic graphs.
For all metrics, node degree, betweenness, clustering, and assortativity,
\ac{\model} substantially reduces the overall dissimilarity concerning the distributions
observed in both training and test sets, as illustrated in Figure~\ref{fig:pdf}.
Despite this significant reduction, none of the generators, including \ac{\model},
accurately replicate the right tails of the distributions for clustering and assortativity
(Figures~\ref{fig:pdfbaclustering}, \ref{fig:pdfnonbaclustering}, \ref{fig:pdfbaassortativity},
and \ref{fig:pdfnonbaassortativity}).
This inability of \ac{\model}
to reproduce these right tails might be attributed to potential overfitting,
given that these right tails exclusively manifest in the test set.
Furthermore, table~\ref{tab:mmd-values} demonstrates that \ac{\model} surpasses
all baselines even when considering the high-order statistical properties assessed through \ac{MMD}.

\section{Conclusions}
\label{sec:conclusion}

This paper introduced \ac{\model}, a novel intra-AS graph generator.
It also introduces 
a novel dataset (\ac{\ds}) composed of real intra-AS graphs. To create \ac{\ds}, we proposed an adaptation of
the multi-level algorithm in~\cite{blondel2008fast} (\ac{\ce}) that is a parameterized algorithm
for subgraphs extraction, which ensures that subgraphs are within predefined limits for a number of nodes
without losing the original characteristics of the graph formation process.

Experimental results demonstrate that the \ac{\model} generator outperforms all baseline generators. On average, \ac{\model} improved the Maximum Mean Discrepancy $(84.4\pm 27.3)\%$, $(95.1\pm 8.9)\%$, $(97.9\pm 3.5)\%$, and $(94.7\pm 9.5)\%$, for assortativity, betweenness, clustering, and node degree, respectively, as shown in Table~\ref{tab:improvements}.
This comparison reveals that graphs generated by \ac{\model} exhibit lower
levels of realism compared to those sampled from the training set, as anticipated, given that the training set is
drawn from the genuine intra-AS graph population (\ac{\ds}). However, despite this disparity, \ac{\model}
demonstrates superior realism when contrasted with the baselines. Consequently,
\ac{\model} exhibits the most notable overall reduction in discrepancy concerning the \ac{MMD} values
of the training samples.

\ac{\ds} is the first dataset that shows such a wide variety of real-world intra-AS graphs and offers novel possibilities for the analysis of 
 solutions for the Internet. \ac{\ds} provides the possibility of training data-driven models based on graphs using
only real-world topologies and improving the generalization capacity of models due to the variety of graphs. 
\ac{\ds} can also be used to diversify the simulation and emulation of solutions for the Internet.

Future work encompasses the investigation of generalization capacity of \ac{\model} and the use of \ac{\model} for graph-based learning algorithms~\cite{tang2021survey,suarez2022graph}. We also plan to develop a user interface for \ac{\model} to help synthesize graphs for intra-AS networks.

\section*{Acknowledgments}
\addcontentsline{toc}{section}{Acknowledgment}
The authors would like to thank CAPES (grant \#88882.329131/2019--01). Authors
are also grateful to CNPq (grant \#307560/2016-3), S\~ao Paulo Research
Foundation -- FAPESP (grants \#2014/12236-1, \#2015/24494-8, and
\#2016/50250-1, \#2017/20945-0).

\printbibliography%

\end{document}